\documentclass[12pt]{article}

\usepackage[dvips]{epsfig}
\usepackage{hyperref}
\usepackage{harvard}       
\usepackage{amsmath, amsthm, amssymb}
\usepackage{xcolor}
\usepackage{multirow}
\usepackage{graphicx}
\usepackage{subfigure}
\usepackage{epstopdf}
\usepackage{cite}
\usepackage{appendix}
\usepackage{listings}
\hypersetup{pdfborder={0 0 0}, colorlinks=false}

\newtheorem{theorem}{Theorem}

\theoremstyle{definition}

\begin{document}
\title{How to estimate the sample mean and standard deviation from the five number summary?}
\author{\small Jiandong Shi$^1$, Dehui Luo$^2$, Hong Weng$^3$, Xian-Tao Zeng$^3$, Lu Lin$^4$ and Tiejun Tong$^2$ \\
\\
{\small $^1$School of Mathematics, Shandong University, China}\\
{\small $^2$Department of Mathematics, Hong Kong Baptist University, Hong Kong}\\
{\small $^3$Center for Evidence-Based and Translational Medicine, Zhongnan Hospital of}\\
{\small Wuhan University, China}\\
{\small $^4$Zhongtai Securities Institute for Finance Studies, Shandong University, China}
}
\date{}
\maketitle

\begin{abstract} 
In some clinical studies, researchers may report the five number summary (including the sample median, the first and third quartiles,
and the minimum and maximum values) rather than the sample mean and standard deviation.
To conduct meta-analysis for pooling studies,
one needs to first estimate the sample mean and standard deviation from the five number summary.
A number of studies have been proposed in the recent literature to solve this problem.
However, none of the existing estimators for the standard deviation is satisfactory for practical use.
After a brief review of the existing literature,
we point out that Wan et al.'s method (BMC Med Res Methodol 14:135, 2014) has a serious limitation in estimating the standard deviation from the five number summary.
To improve it, we propose a smoothly weighted estimator by incorporating the sample size information and derive the optimal weight for the new estimator.
For ease of implementation, we also provide an approximation formula of the optimal weight and a shortcut formula for estimating the standard deviation from the five number summary.
The performance of the proposed estimator is evaluated through two simulation studies.
In comparison with Wan et al.'s estimator, our new estimator provides a more accurate estimate for normal data and performs favorably for non-normal data.
In real data analysis, our new method is also able to provide a more accurate estimate of the true sample standard deviation than the existing method.
In this paper, we propose an optimal estimator of the standard deviation from the five number summary.
Together with the optimal mean estimator in Luo et al. (Stat Methods Med Res, in press, 2017),
our new methods have improved the existing literature and will make a solid contribution to meta-analysis and evidence-based medicine.

\end{abstract}

\newpage
\baselineskip 18pt
\setlength{\parskip}{0.2\baselineskip}
\section{Introduction}
Meta-analysis is becoming increasingly popular during the past two decades, mainly due to its wide applications in evidence-based medicine.
To statistically combine data from multiple studies using meta-analysis, one usually needs to conduct a systematic review and extract the summary data from the clinical studies in the literature.
For continuous outcomes, e.g., the high blood pressure and the amount of alcohol consumed,
the sample mean and standard deviation are two commonly used statistics reported for evaluating the effectiveness of a certain medicine or treatment.
We also note that, for certain reasons, the sample median, the first and third quartiles, and/or the minimum and maximum values are also frequently reported in the literature.
To the best of our knowledge, no existing methods in meta-analysis can handle the sample median and the sample mean simultaneously.
As one example, when applying the fixed-effect model or the random-effects model, we are allowed to use only the sample mean and standard deviation data to derive the overall effect size of the treatment.

Then, a natural question is: how to deal with the studies in the literature with the sample median data?
In the early stages, researchers often exclude such studies from further analysis by claiming them as ``Studies with no sufficient data" in the flow chart of study selection.
Such an approach, however, is suboptimal as it may lose valuable information in the literature.
As a consequence, the final results are usually less reliable,
in particular when the total number of studies is small or a large proportion of those studies are reported with the sample median data.
For this, there is an increased demand for developing new methods that convert the sample median data to the sample mean data for meta-analysis.
For ease of notation, let $\{a,q_1,m,q_3,b\}$ denote the five number summary, where $a$ is the sample minimum,
$q_1$ is the first quartile, $m$ is the sample median, $q_3$ is the third quartile, and $b$ is the sample maximum of the data.
Let also $n$ be the sample size of the study.

Note that the five number summary may not be fully reported in clinical studies.
In the special case where only $\{a,m,b\}$ were reported, Hozo et al. \cite{hozo2005estimating} was the first to provide a simple method for estimating the sample mean and standard deviation.
It is noted, however, that Hozo et al. \cite{hozo2005estimating} did not sufficiently use the information of sample size $n$ so that their estimators are either biased or non-smooth.
Inspired by this, to improve Hozo et al. \cite{hozo2005estimating}'s method,
Wan et al. \cite{wan2014estimating} proposed an unbiased estimator of the standard deviation,
and Luo et al. \cite{luo2016optimally} proposed an optimal estimation of the sample mean.
In another special case where only $\{q_1, m, q_3\}$ were reported,
Wan et al. \cite{wan2014estimating} also proposed a simple estimator of the sample mean and an unbiased estimator of the standard deviation.
Luo et al. \cite{luo2016optimally} further improved the sample mean estimator in Wan et al. \cite{wan2014estimating} by providing the optimal weights between the two components of the estimator using the sample size information.
In Google Scholar on 10 September 2017, Hozo et al. \cite{hozo2005estimating} has been cited 1936 times and Wan et al. \cite{wan2014estimating} has been cited 203 times.
Without any doubt, these several papers have been attracting more attentions and playing an important role in meta-analysis.

When $\{a,q_1,m,q_3,b\}$ were fully reported, Bland \cite{Bland2015Estimating} proposed some new estimators for the sample mean and standard deviation from the five number summary.
As his method is essentially the same as Hozo et al.'s method, Bland's estimators are also suboptimal and the sample size information is not sufficiently used.
For instance, the sample mean estimator in Bland \cite{Bland2015Estimating} is
\begin{equation}\label{blandmean}
  \bar{X}\approx\dfrac{a+2q_1+2m+2q_3+b}{8}=\dfrac{1}{4}\left(\dfrac{a+b}{2}\right)+\dfrac{1}{2}\left(\dfrac{q_1+q_3}{2}\right)+\dfrac{m}{4}.
\end{equation}
Given that the data follow a symmetric distribution, the quantities $(a+b)/2$, $(q_1+q_3)/2$, and $m$ can each serve as an estimate of the sample mean.
To have a final estimator, Bland \cite{Bland2015Estimating} applied the artificial weights 1/4, 1/2, and 1/4 for the three components, respectively.
That is, the first and third components are treated equally and both of them are only half reliable compared to the second component.
As this is not always the truth, to improve the sample mean estimation, Luo et al. \cite{luo2016optimally} proposed the optimal estimator as
\begin{equation}
\bar{X}\approx w_1\left(\dfrac{a+b}{2}\right)+w_2\left(\dfrac{q_1+q_3}{2}\right)+(1-w_1-w_2)m,
\label{mabq1q3}
\end{equation} 
\noindent where $w_1=2.2/(2.2+n^{0.75})$ and $w_2=0.7-0.72/n^{0.55}$ are the optimal weights assigned to the respective components.

For the standard deviation estimation from the five number summary,
Bland \cite{Bland2015Estimating} also provided an estimator by the inequality method as in Hozo et al. (see also estimator (\ref{bland}) in Section 2).
Wan et al. \cite{wan2014estimating} then proposed an unbiased estimator of the standard deviation to improve Bland's method.
In Section 3, however, we will point out that the standard deviation estimator in Wan et al. \cite{wan2014estimating} is still suboptimal due to the insufficient use of the sample size information.
For more details, see the motivation example in Section 3.
According to Higgins \& Green \cite{higgins2011cochrane} and Chen \& Peace \cite{chen2013applied}, the sample standard deviations play a crucial role in weighting the studies in meta-analysis.
Inaccurate weighting results may lead to biased overall effect sizes and biased confidence intervals, and hence mislead physicians to provide patients with unreasonable or even wrong medications.
Inspired by this, we propose a smoothly weighted estimator of the standard deviation to improve the existing methods.
From the practical point of view, our proposed method will make a solid contribution to meta-analysis.

The rest of the paper is organized as follows.
In Section 2, we review the existing methods of estimating the standard deviation under three common scenarios.
In Section 3, we present a motivation example to explore the limitations of the current method,
propose a smoothly weighted estimator of the standard deviation, and also derive some analytical results to support our new method.
We then conduct simulation studies and a real data analysis in Section 4 to demonstrate the advantages of our new estimator of the standard deviation.
Finally, we discuss some interesting problems in Section 5 and conclude the paper in Section 6.

\section{Existing methods}
Needless to say, the sample size $n$ is also an important information and should be sufficiently used in the estimation procedure.
To include it with the five number summary, we let $\mathcal{S}_1=\{a, m, b; n\}$, $\mathcal{S}_2=\{q_1, m, q_3; n\}$,
and $\mathcal{S}_3=\{a, q_1, m, q_3, b; n\}$ to represent the three scenarios frequently appeared in the literature.
It is clear that $\mathcal{S}_1$ and $\mathcal{S}_2$ are two special cases of $\mathcal{S}_3$.
To avoid confusion, we also note that $\mathcal{S}_1$, $\mathcal{S}_2$ and $\mathcal{S}_3$ are the same as $\mathcal{C}_1$, $\mathcal{C}_3$ and $\mathcal{C}_2$ in Wan et al. \cite{wan2014estimating}, respectively.
In this section, we briefly review the existing estimators of the standard deviation under the three scenarios.

\subsection{Estimating the standard deviation from $\mathcal{S}_1 = \{a, m, b; n\}$}
For scenario $\mathcal{S}_1$, Hozo et al. \cite{hozo2005estimating} proposed to estimate the standard deviation by
$$
S\approx\left\{
\begin{aligned}
&  \dfrac{1}{\sqrt{12}}\left[(b-a)^2+\dfrac{(a-2m+b)^2}{4}\right]^{1/2} \quad & n\leq15,\\
& \dfrac{b-a}{4}  & 15<n\leq70,\\
& \dfrac{b-a}{6}  &n>70.
\end{aligned}
\right.
$$
\noindent As shown in Google Scholar, Hozo et al.'s estimator is very popular in the previous literature due to the huge demand of data transformation in evidence-based medicine.
We note, however, that their estimator is a step function of the sample size $n$ so that the final estimate may not be reliable.
For example, when $n$ increases from 70 to 71, there is a sudden drop in the estimated value of the standard deviation,
namely dropped by 33.3\% if the 71st sample is not an  extreme value so that $a$ and $b$ remain the same in both samples.
On the other hand, the sample size information is completely ignored within each of the three intervals so that their estimator is biased and suboptimal.

In view of the above limitations, Wan et al. \cite{wan2014estimating} proposed an unbiased estimator of the standard deviation as
\begin{equation}
  S\approx\dfrac{b-a}{\xi},
  \label{ab}
\end{equation}
where $\xi=\xi(n)=2\Phi^{-1}[(n-0.375)/(n+0.25)]$, $\Phi$ is the cumulative distribution function of the standard normal distribution, and $\Phi^{-1}$ is the inverse function of $\Phi$.
Wan et al.'s estimator not only overcomes the limitations of Hozo et al.'s estimator by incorporating the sample size information efficiently, but also is very simple for practical use.
From a statistical point of view, Wan et al. \cite{wan2014estimating} has provided the best method for estimating the standard deviation,
given that the reported data include only the four numbers in scenario $\mathcal{S}_1$.

\subsection{Estimating the standard deviation from $\mathcal{S}_2 = \{q_1, m, q_3; n\}$}
For scenario $\mathcal{S}_2$, Wan et al. \cite{wan2014estimating} proposed an estimator of the standard deviation as
\begin{equation}
  S\approx\dfrac{q_3-q_1}{\eta},
  \label{q1q3}
\end{equation}
where $\eta=\eta(n)=2\Phi^{-1}[(0.75n-0.125)/(n+0.25)]$.
Note that the method for deriving estimator (\ref{q1q3}) is the same in spirit as that for estimator (\ref{ab}).
In particular, Wan et al. have incorporated the sample size information appropriately so that the proposed estimator is approximately unbiased.
When the reported data include only $q_1$, $m$, $q_3$ and $n$, it can be shown that estimator (\ref{q1q3}) is the best method for estimating the standard deviation.

\subsection{Estimating the standard deviation from $\mathcal{S}_3 = \{a, q_1, m, q_3, b; n\}$}
For scenario $\mathcal{S}_3$, Bland \cite{Bland2015Estimating} proposed to estimate the standard deviation by
\begin{equation}\label{bland}
\begin{split}
 S \approx &  \ \left[{(a^2+2q_1^2+2m^2+2q_3^2+b^2) \over 16}+{(aq_1+q_1m+mq_3+q_3b)\over 8} \right.\\
           &  \ \left. -{(a+2q_1+2m+2q_3+b)^2\over 64} \large\right]^{1/2}.
\end{split}
\end{equation}
\noindent 
As mentioned in Section 1, Bland's estimator is independent of the sample size and is less accurate for practical use.
To improve the literature, Wan et al. \cite{wan2014estimating} proposed the following estimator of the standard deviation:
\begin{equation}\label{abq}
  S\approx\dfrac{1}{2}\left(\dfrac{b-a}{\xi}+\dfrac{q_3-q_1}{\eta}\right).
\end{equation}

In essence, Wan et al. \cite{wan2014estimating} treated scenario $\mathcal{S}_3$ as a combination of scenario $\mathcal{S}_1$ and scenario $\mathcal{S}_2$.
By estimating the standard deviation from scenario $\mathcal{S}_1$ and scenario $\mathcal{S}_2$ separately, they applied the average of estimators (\ref{ab}) and (\ref{q1q3}) as the final estimator.
In Section 3, we will show that $(b-a)/\xi$ and $(q_3-q_1)/\eta$ may not be equally reliable for different sample size $n$.
As a consequence, the average estimator in (\ref{abq}) is not the optimal estimator due to the insufficient use of the sample size information.
This motivates us to propose a smoothly weighted estimator of the standard deviation to further improve Wan et al.'s estimator in this paper.

\section{Main results}
To present the main idea, we let $X_{1} , X_{2} , \ldots , X_{n}$ be a random sample of size $n$ from the normal distribution with mean $\mu$ and variance $\sigma^2$,
and $X_{(1)} \leq X_{(2)} \leq \cdots \leq X_{(n)}$ be the order statistics of $X_{1} , X_{2} , \ldots , X_{n}$.
Let also $X_i = \mu+\sigma Z_i$ for $i=1, 2, \ldots,n$.
Then $Z_{1} , Z_{2} , \ldots , Z_{n}$ follow the standard normal distribution with $Z_{(1)}\leq Z_{(2)} \leq \cdots \leq Z_{(n)}$ as the corresponding order statistics.
Finally, by letting $n = 4Q+1$ with $Q$ a positive integer, we have $a = \mu+\sigma Z_{(1)}$, $q_1 = \mu+\sigma Z_{(Q+1)}$, $m = \mu+\sigma Z_{(2Q+1)}$, $q_3 = \mu+\sigma Z_{(3Q+1)}$, and $b = \mu+\sigma Z_{(n)}$.

\subsection{Motivation example}
To investigate whether the two components in estimator (\ref{abq}) are equally reliable, we first conduct a simple simulation study.
In each simulation, we generate a random sample of size $n$ from the standard normal distribution, find the five number summary $\{a,q_1,m,q_3,b\}$,
and then apply estimators (\ref{ab}) and (\ref{q1q3}) to estimate the standard deviation respectively.
For $n=5$ and $n=5001$, we repeat the above simulation 10,000 times and plot the histograms of the simulated estimates for both methods in Figure 1.

\begin{figure}[!hbt]
  \centering
  \includegraphics[width=0.9\textwidth]{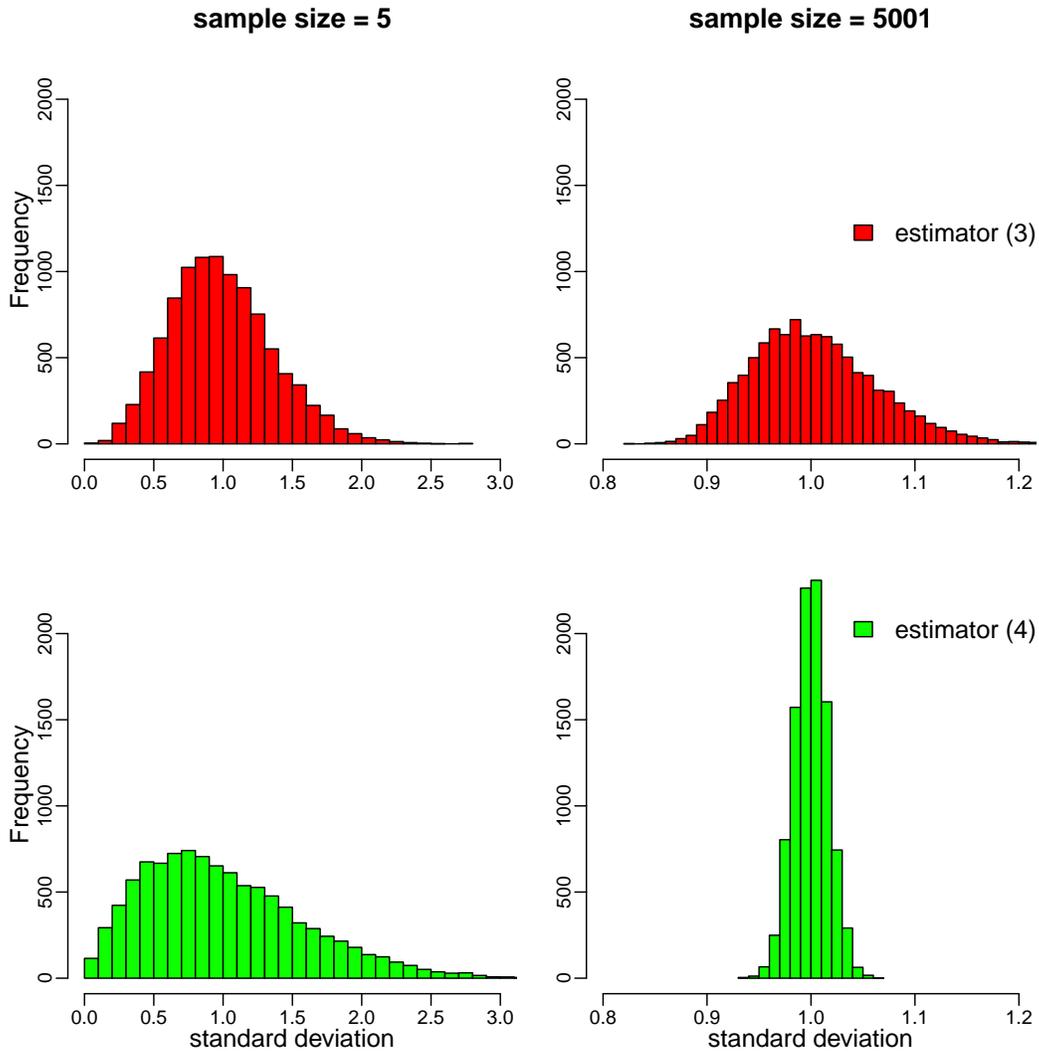}
  \caption{Histograms of the simulated standard deviations (the true value is 1) for various settings.}\label{fig:motivation}
\end{figure}

From Figure 1, it is evident that estimators (\ref{ab}) and (\ref{q1q3}) are not equally reliable for different sample sizes.
Note that the true value of the standard deviation is 1 as the data are simulated from the standard normal distribution.
When the sample size is small (say $n=5$), estimator (\ref{ab}) provides a more accurate and less skewed estimate for the standard deviation.
Instead, when the sample size is large (say $n=5001$), estimator (\ref{q1q3}) is a much better estimator than estimator (\ref{ab}).
Inspired by this, we propose to further improve Wan et al.'s estimator by considering a weighted combination of estimators (\ref{ab}) and (\ref{q1q3}), in which the optimal weight will be determined by the sample size.

\subsection{Optimal standard deviation estimation}
In view of the limitations of estimator (\ref{abq}), we propose the following estimator for the standard deviation:
\begin{equation}\label{abq1q3}
S(w)=w\left(\dfrac{b-a}{\xi}\right)+(1-w)\left(\dfrac{q_{3}-q_{1}}{\eta}\right),
\end{equation}
where $\xi$ and $\eta$ are given in Section 2,
and $w$ is the weight assigned to the first component.
Note that the new estimator is a weighted combination of estimators (\ref{ab}) and (\ref{q1q3}).
When $w=1$, the new estimator reduces to estimator (\ref{ab}).
When $w=0$, the new estimator reduces to estimator (\ref{q1q3}).
And when $w=0.5$, the new estimator leads to estimator (\ref{abq}) in Wan et al. \cite{wan2014estimating}.
Hence, the existing estimators of the standard deviation are all special cases of our new estimator.

To find the optimal estimator, we consider the commonly used quadratic loss function, i.e., $L(S(w), \sigma) = (S(w)-\sigma)^2$.
We select the optimal weight by minimizing the expectation of the loss function $L(S(w),\sigma)$, or equivalently, by minimizing the mean squared error (MSE) of the estimator.
In Theorem 1 of the Appendix, we show that the optimal weight has the following form:
\begin{equation}\label{weight}
w_{\rm opt}=\dfrac{{\rm Var}(q_{3}-q_{1})/\eta^{2}-{\rm Cov}(b-a, q_{3}-q_{1})/(\xi\eta)}{{\rm Var}(b-a)/\xi^{2}+{\rm Var}(q_{3}-q_{1})/\eta^{2}-2{\rm Cov}(b-a,q_{3}-q_{1})/(\xi\eta)}.
\end{equation}
Apparently, the analytical form of the optimal weight is quite complicated for practitioners.
For practical use, we propose to simplify the analytical form by developing an approximation formula for the optimal weight.
Recall that $a = \mu+\sigma Z_{(1)}$, $q_1 = \mu+\sigma Z_{(Q+1)}$, $q_3 = \mu+\sigma Z_{(3Q+1)}$, and $b = \mu+\sigma Z_{(n)}$.
We have $b-a=\sigma(Z_{(n)}-Z_{(1)})$ and $q_3-q_1 = \sigma(Z_{(3Q+1)}-Z_{(Q+1)})$.
Then by formula (8) and the symmetry of the standard normal distribution, we can rewrite the optimal weight as
\begin{equation}\label{approximation}
w_{\rm opt}(n)=\dfrac{1}{1+J(n)},
\end{equation}
where
\begin{equation}\label{Jn}
J(n)=\dfrac{{\rm Var}(Z_{(n)}-Z_{(1)})/\xi^2-{\rm Cov}(Z_{(n)}-Z_{(1)})/(\xi\eta)}{{\rm Var}(Z_{(3Q+1)}-Z_{(Q+1)})/\eta^2-{\rm Cov}(Z_{(n)}-Z_{(1)})/(\xi\eta)}.
\end{equation}
Note that $J(n)$ is independent of the parameters $\mu$ and $\sigma^2$ and depends only on the sample size $n$.
This shows that the optimal weight $w_{\rm opt}$ is a function of $n$ only.
For clarification, we have expressed the optimal weight as $w_{\rm opt} = w_{\rm opt}(n)$ in (\ref{approximation}).

\subsection{An approximation formula}
To have an approximation formula for the optimal weight, we numerically compute the true values of $J(n)$ and $w_{\rm opt}(n)$ for different values of $n$ using formulas (\ref{approximation}) and (\ref{Jn}).
We then plot $J(n)$ and $w_{\rm opt}(n)$ in Figure 2 for $n$ varying from 5 to 401, respectively.
Observing that $J(n)$ is an increasing and concave function of $n$, we consider the simple power function $c_1n^{c_2}+c_0$ to approximate $J(n)$ with $0<c_2<1$ so that the approximation curve is also concave.
With the true values of $J(n)$, the best values of the coefficients are approximately $c_1=0.07$, $c_2=0.6$ and $c_0=0$; that is, we propose to use $0.07n^{0.6}$ to approximate $J(n)$.
Next, by plugging $J(n)\approx0.07n^{0.6}$ into (\ref{approximation}), we have the approximated optimal weight as
\begin{equation}\label{aweight}
\tilde{w}_{\rm opt}(n)\approx \dfrac{1}{1+0.07n^{0.6}}.
\end{equation}

\begin{figure}[!hbt]
  \centering
  \includegraphics[width=0.9\textwidth]{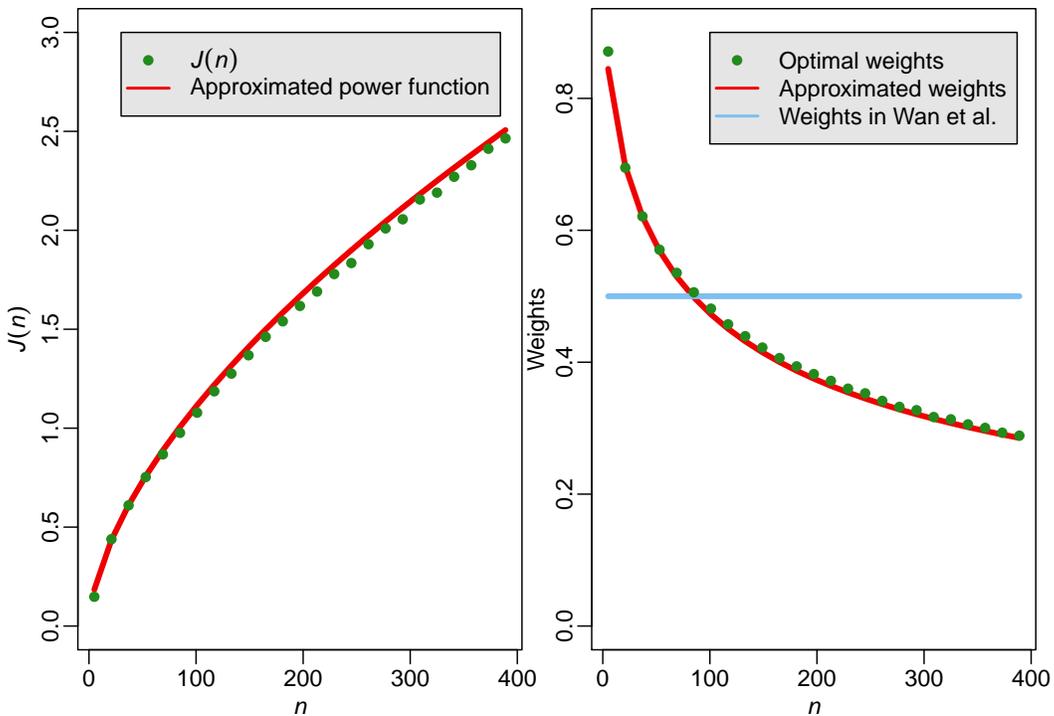}
  \caption{The true values and approximated formula of $J(n)$ are shown on the left, and the corresponding weights are shown on the right.}\label{fig:optimalweight}
\end{figure}

\begin{figure}[!hbt]
  \centering
  \includegraphics[width=0.9\textwidth]{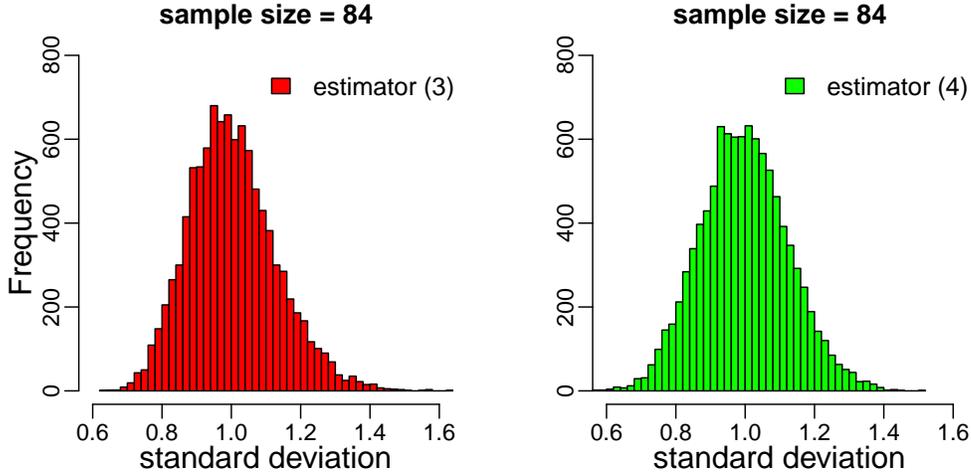}
  \caption{Histograms of the simulated standard deviations (the true value is 1) when $n=84$.}\label{fig:demonstration}
\end{figure}

From the right panel of Figure 2, it is evident that the approximation formula provides a perfect fit to the true optimal weight values for $n$ up to $401$.
By (\ref{abq1q3}) and (\ref{aweight}), our proposed estimator of the standard deviation from the five number summary is
\begin{equation}\label{approx}
S(n)\approx\left(\dfrac{1}{1+0.07n^{0.6}}\right)\dfrac{b-a}{\xi}+\left(\dfrac{0.07n^{0.6}}{1+0.07n^{0.6}}\right)\dfrac{q_3-q_1}{\eta}.
\end{equation}

\noindent From the approximation formula (\ref{aweight}), $n$ is about 84 so that $\tilde{w}_{\rm opt}(n)\approx0.5$.
That is, our approximation formula indicates that when $n=84$, estimators (\ref{ab}) and (\ref{q1q3}) will be equally reliable.
To examine and verify it, we conduct another simulation study with $n=84$.
All other settings are kept the same as in the motivation example in Section 3.1.
From the histograms in Figure 3, we note that estimators (\ref{ab}) and (\ref{q1q3}) perform very similarly when $n=84$.
This, from another perspective, demonstrates that our approximation formula can serve as a ``rule of thumb'' for estimating the standard deviation from the five number summary.

Recall that $\xi=\xi(n)=2\Phi^{-1}[(n-0.375)/(n+0.25)]$ and $\eta=\eta(n)=2\Phi^{-1}[(0.75n-0.125)/(n+0.25)]$,
where $\Phi^{-1}(z)$ is the upper $z$th percentile of the standard normal distribution.
We have the shortcut formula of (\ref{approx}) as
\begin{equation}\label{shortcut}
S(n)\approx\dfrac{b-a}{\theta_1(n)}+\dfrac{q_3-q_1}{\theta_2(n)},
\end{equation}
\noindent where
\begin{align*}
\theta_1(n) &= (2+0.14n^{0.6})\cdot \Phi^{-1}\left(\frac{n-0.375}{n+0.25}\right),\\
\theta_2(n) &= \left(2+\frac{2}{0.07n^{0.6}}\right)\cdot \Phi^{-1}\left(\frac{0.75n-0.125}{n+0.25}\right).
\end{align*}
For ease of implementation, we also provide the numerical values of $\theta_1(n)$ and $\theta_2(n)$ in Table \ref{coefficients} for $Q$ up to 60, or equivalently, for $n$ up to 241.
For a general sample size, one may refer to our Excel spread sheet for specific values, or compute them using the command ``qnorm($z$)" in the R software.

\section{Data analysis}
\subsection{Simulation study}
To evaluate the performance of the proposed method, we conduct two simulation studies to compare our new estimator with the existing one in Wan et al. \cite{wan2014estimating}.
The robustness of the two estimators will also be examined.

In the first study, we generate the data from the normal distribution with mean $\mu=50$ and standard deviation $\sigma=17$,
which follows from Hozo et al. \cite{hozo2005estimating} and Wan et al. \cite{wan2014estimating} for a fair comparison.
For each generated sample of size $n$, we compute the sample standard deviation $S$ and also record the five number summary $\{a , q_1, m, q_3, b \}$.
To apply the proposed method, we presume that the full sample is not reported and the only available data are the five number summary and the sample size.
We then use estimators (\ref{abq}) and (\ref{shortcut}) to get the estimates of the standard deviation, and denote them by $S^{\rm Ex}$ and $S^{\rm New}$ respectively.
To compare them, we consider the relative mean squared error (RMSE) as the criterion:
$${\rm RMSE}(S^{\rm Ex})=\dfrac{\sum_{i=1}^{T}(S^{\rm Ex}_i-\sigma)^2}{\sum_{i=1}^{T}(S_i-\sigma)^2}  ~~~~~ {\rm and} ~~~~~  {\rm RMSE}(S^{\rm New})=\dfrac{\sum_{i=1}^{T}(S^{\rm New}_i-\sigma)^2}{\sum_{i=1}^{T}(S_i-\sigma)^2},$$
where $T$ is the total number of repetitions, $\sigma$ is the true standard deviation, and $S_i$ is the sample standard deviation of the $i$th sample.

\begin{figure}[!htb]
  \centering
  \includegraphics[width=0.75\textwidth]{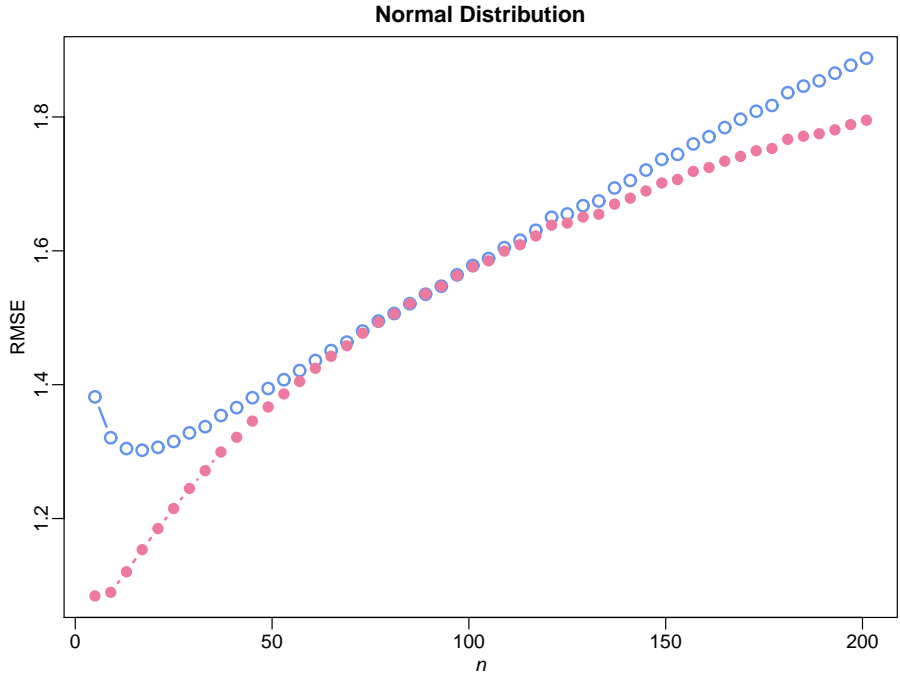}
  \caption{RMSE values of the standard deviation estimators for data from the normal distribution. The empty points represent Wan et al.'s method, and the solid points represent our new method.}\label{fig:norm}
\end{figure}

With a total of $T=2,000,000$ repetitions, we report the RMSE values of the two estimators in Figure 4 with $n$ ranging from 5 to 201.
From the simulation results, it is evident that our new estimator always has a smaller RMSE value than Wan et al.'s estimator, in particular when $n$ is very small or very large.
When $n$ is close to 84, the two estimators have a similar performance.
This coincides with the analytical result in Section 3 that the optimal weight $w_{\rm opt}(n)$ approaches 0.5 when $n$ is about 84.
In conclusion, our new estimator performs consistently better than Wan et al.'s estimator when the data are normal.

\begin{figure}[!htb]
  \centering
  \includegraphics[width=0.9\textwidth]{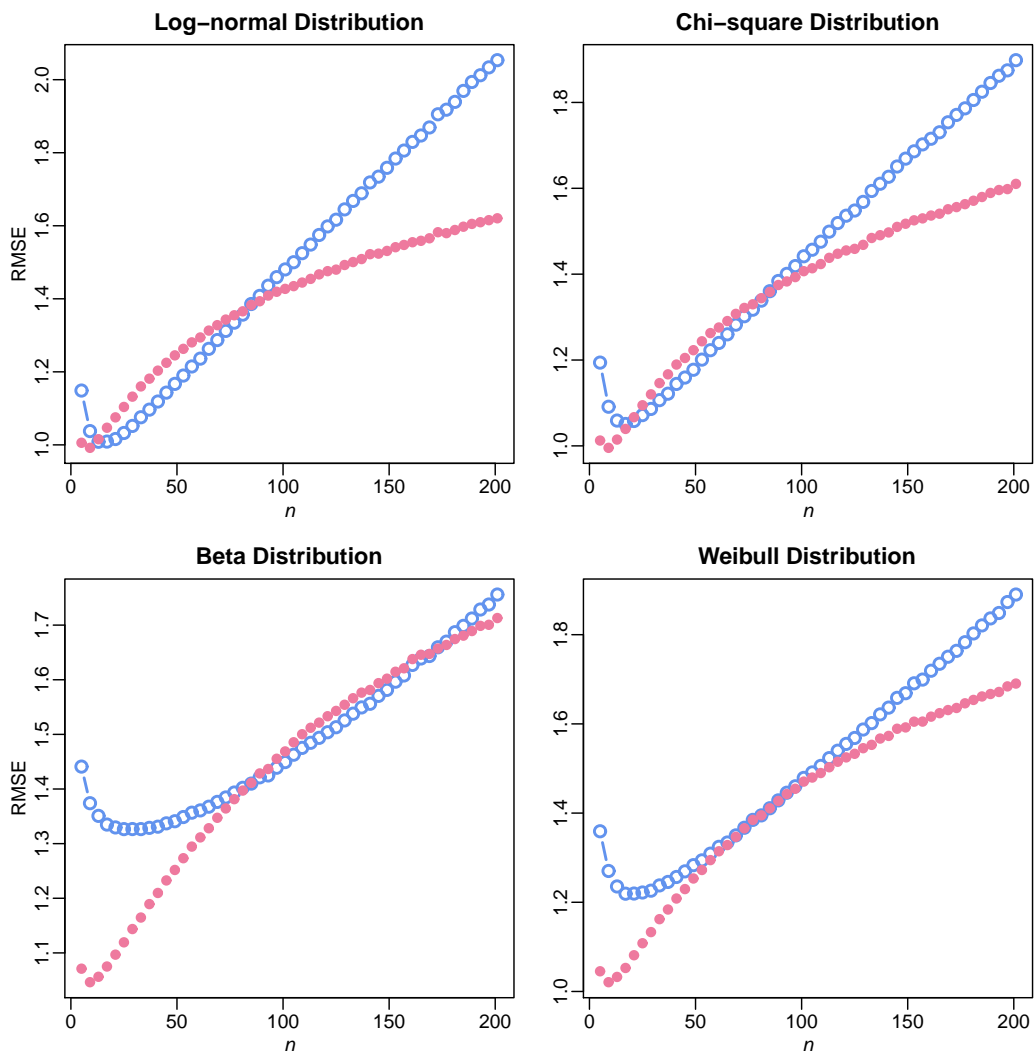}
  \caption{RMSE values of the standard deviation estimators for data from the skewed distributions. The empty points represent Wan et al.'s method, and the solid points represent our new method.}\label{fig:distribution}
\end{figure}

To examine the robustness of the proposed method, we conduct another simulation study in which the data are generated from non-normal distributions.
Specifically, we consider four skewed distributions including the log-normal distribution with location parameter $\mu=4$ and scale parameter $\sigma=0.3$, the chi-square distribution with 10 degrees of freedom,
the beta distribution with shape parameters $\alpha=9$ and $\beta=4$, and the Weibull distribution with shape parameter $k=2$ and scale parameter $\lambda=35$.
Other settings and the estimation procedure follow the same as in the first study.
Then for estimators (\ref{abq}) and (\ref{shortcut}), we also compute the RMSE values and use them to evaluate and compare the robustness of the two methods.

With $500,000$ simulations for each setting, we report their respective RMSE values in Figure 4 for $n$ up to 201.
Overall, our new estimator provides a smaller RMSE value than Wan et al.'s estimator in most settings regardless of the level of skewness.
We also note that the RMSE of our new estimator increases more slowly as the sample size increases.
This implies that our new method has provided an asymptotically more efficient estimator even when the data are not symmetric.
Together with the comparison results in the first study, we conclude that our new estimator not only provides a more accurate estimate for normal data,
but also performs favorably compared to the existing method for non-normal data.

\subsection{Real data analysis}
To illustrate the usefulness of the new method, we consider a clinical study in which the range,
the interquartile range (IQR) and the sample standard deviation (sample SD) of the data were all reported.
For rodents, some studies suggested that the $n-3$ long-chain polyunsaturated fatty acids ($n-3$ PUFA) as peroxisome proliferator-activated receptor-$\alpha$ ligands may improve the non-alcoholic fatty liver disease (NAFLD).
In contrast, there were few studies reported in the literature with human beings as subjects.
To evaluate the efficacy of prolonged PUFA supplementation in patients with NAFLD,
Capanni et al. \cite{capanni2006prolonged} conducted a clinical trial with 42 patients accepting the treatment as the PUFA group and 14 patients refusing the treatment as the control group.
Body mass index (BMI) and doppler perfusion index (DPI) were adopted to evaluate the condition of the patients.
Data of the two indexes are reported in Table \ref{summary}.

With the reported data in their paper, we use estimators (\ref{abq}) and (\ref{shortcut}) to estimate the standard deviation respectively,
and report those estimates in Table \ref{summary} as well for comparison.
Treating the sample SD as the true values, our new estimator always provides a more accurate estimate than Wan et al.'s estimator.
In particular for the treatment group with $n=42$, the estimation error of our new estimator is less than half of that for Wan et al.'s estimator for both BMI and DPI.

\begin{table}[htbp] \scriptsize
\caption{\small Summary data from the study of Capanni et al.} \label{summary}
\centering
\begin{tabular}{ccccccc}
  \hline
  Sample size              &Index               & range               & IQR               & sample SD              & Wan et al.'s estimates             & Our estimates\\   \hline
   \multirow{2}{*}{14}      &BMI                 & 22.8-34.3           & 4                 & 4                           & 3.331                               & 3.348\\
                            &DPI                 & 0.04-0.19           & 0.04              & 0.04                        & 0.038                               & 0.041\\         \hline
   \multirow{2}{*}{42}      &BMI                 & 23-38.6             & 8.1               & 4.7                         & 4.901                               & 4.631\\
                            &DPI                 & 0.06-0.24           & 0.09              & 0.05                        & 0.055                               & 0.052\\
  \hline
\end{tabular}
\end{table}

\setlength{\parskip}{0.2\baselineskip}
\section{Discussion}
In clinical studies with continuous outcomes, the sample mean and standard deviation are routinely reported;
while in some studies, one may also report the five number summary including the sample median, the first and third quartiles, and the minimum and maximum values.
To conduct meta-analysis for pooling studies in the literature, researchers need to first transform the five number summary to the sample mean and standard deviation.
As reviewed in Section 2, a number of studies have been proposed recently to solve this problem under several common scenarios.
It is noted, however, that there is a serious limitation in estimating the standard deviation from the five number summary in Wan et al. \cite{wan2014estimating}.
Inspired by this, we have proposed an improved estimator of the standard deviation by considering a smoothly weighted estimator and further derived the analytical form of the optimal weight.
Given that the analytical form is quite complicated and may not be accessible for practitioners,
we have also derived an approximation formula for the optimal weight.
Our shortcut formula (\ref{shortcut}) can serve as a ``rule of thumb" for estimating the standard deviation from the five number summary.

To our understanding, most existing methods for estimating the sample mean and standard deviation from the five number summary are established on the basis of the assumption that the data are normally distributed.
Note, however, that the real data from clinical studies may not follow a normal distribution.
For non-normal data, the estimation accuracy of the proposed methods including the one in our paper may not always be guaranteed,
in particular when the data are relatively skewed.
As a future work, we will develop some robust methods for estimating the sample mean and standard deviation from non-normal data,
no matter whether or not the underlying distribution is specified.
To promote the practical use of our new method, we have also provided an Excel spread sheet including the optimal mean estimator (\ref{mabq1q3}) and the optimal standard deviation estimator (\ref{shortcut}).
To be specific, when the reported data are the five number summary,
one can simply input those data and the sample size into the designated places in the Excel file.
The estimates of the sample mean and standard deviation will then be displayed automatically.
For comparison, the standard deviation estimator (\ref{abq}) in Wan et al. \cite{wan2014estimating} is also included in the spread sheet.

\section{Conclusions}
In this paper, we proposed a smoothly weighted estimator of the standard deviation to improve the literature of meta-analysis,
given that the five number summary were reported in the studies.
We also provided an approximation formula of the optimal weight and a shortcut formula for ease of implementation.
Together with the optimal mean estimator in Luo et al. \cite{luo2016optimally},
our proposed ``rules of  thumb" for estimating the sample mean and standard deviation from the five number summary are as follows:
\begin{equation}\nonumber
\bar{X}\approx w_1\left(\dfrac{a+b}{2}\right)+w_2\left(\dfrac{q_1+q_3}{2}\right)+(1-w_1-w_2)m
\end{equation}
and
\begin{equation}\nonumber
S\approx\dfrac{b-a}{\theta_1}+\dfrac{q_3-q_1}{\theta_2},
\end{equation}
where $w_1=2.2/(2.2+n^{0.75})$, $w_2=0.7-0.72/n^{0.55}$, $\theta_1=(2+0.14n^{0.6})\cdot \Phi^{-1}\left((n-0.375)/(n+0.25)\right)$,
and $\theta_2=\left(2+2/(0.07n^{0.6})\right)\cdot \Phi^{-1}((0.75n-0.125)/(n+0.25))$ with $n$ the sample size of the study.
Simulation studies and a real data analysis were also provided to demonstrate that our new methods have improved the existing literature and will make a solid contribution to meta-analysis and evidence-based medicine.

\section*{Additional files}\small
\textbf{Additional file 1: An Excel spread sheet for estimating the sample mean and standard deviation from the five number summary. }

\noindent \textbf{Additional file 2: Analytical results of the optimal estimator and their derivation.}

\bibliographystyle{plain}
\bibliography{myref}

\begin{table}[!hbt]\normalsize
\caption{Values of $\theta_1(n)$ and $\theta_2(n)$ in formula (\ref{shortcut}) for $1\leq Q\leq60$, where $n=4Q+1$.}\label{coefficients}
\centering
\begin{tabular}{ccc|ccc|ccc}
\hline
$Q$ &$\theta_1(n)$	&$\theta_2(n)$~~  &$Q$	   &$\theta_1(n)$	&$\theta_2(n)$	 ~~&$Q$	&$\theta_1(n)$	   &$\theta_2(n)$\\
\hline
1	&2.7933	&6.4030 ~~&21	&9.7934	    &2.6436 ~~&41	&13.3453	&2.2295\\
2	&3.7701	&5.5135	~~&22	&9.9980	    &2.6098	~~&42	&13.5016	&2.2173\\
3	&4.4372	&4.8952	~~&23	&10.1987	&2.5782	~~&43	&13.6564	&2.2054\\
4	&4.9688	&4.4657	~~&24	&10.3957	&2.5487	~~&44	&13.8099	&2.1940\\
5	&5.4227	&4.1497	~~&25	&10.5893	&2.5210	~~&45	&13.9619	&2.1830\\
6	&5.8255	&3.9063	~~&26	&10.7796	&2.4949	~~&46	&14.1127	&2.1724\\
7	&6.1919	&3.7120	~~&27	&10.9669	&2.4703	~~&47	&14.2622	&2.1621\\
8	&6.5307	&3.5525	~~&28	&11.1513	&2.4471	~~&48	&14.4104	&2.1522\\
9	&6.8478	&3.4189	~~&29	&11.3330	&2.4251	~~&49	&14.5575	&2.1426\\
10	&7.1472	&3.3049	~~&30	&11.5122	&2.4043	~~&50	&14.7034	&2.1333\\
11	&7.4320	&3.2063	~~&31	&11.6890	&2.3845	~~&51	&14.8482	&2.1242\\
12	&7.7043	&3.1199	~~&32	&11.8634	&2.3657	~~&52	&14.9919	&2.1155\\
13	&7.9659	&3.0435	~~&33	&12.0357	&2.3478	~~&53	&15.1346	&2.1070\\
14	&8.2181	&2.9753	~~&34	&12.2058	&2.3307	~~&54	&15.2762	&2.0987\\
15	&8.4621	&2.9140	~~&35	&12.3740	&2.3143	~~&55	&15.4169	&2.0907\\
16	&8.6987	&2.8585	~~&36	&12.5402	&2.2987	~~&56	&15.5565	&2.0829\\
17	&8.9286	&2.8080	~~&37	&12.7046	&2.2837	~~&57	&15.6952	&2.0753\\
18	&9.1526	&2.7618	~~&38	&12.8673	&2.2693	~~&58	&15.8330	&2.0680\\
19	&9.3710	&2.7193	~~&39	&13.0282	&2.2555	~~&59	&15.9699	&2.0608\\
20	&9.5845	&2.6800	~~&40	&13.1876	&2.2423	~~&60	&16.1059	&2.0538\\
\hline
\end{tabular}
\end{table}

\newpage
\begin{appendices}
\section{Appendix} \vskip 12pt
\begin{theorem}
For the weighted estimator of the standard deviation in (\ref{abq1q3}),
$$S(w)=w\left(\dfrac{b-a}{\xi}\right)+(1-w)\left(\dfrac{q_{3}-q_{1}}{\eta}\right),$$
we have the following conclusions:\\
\begin{enumerate}
\item[{\rm (i)}] $S(w)$ is an unbiased estimator of $\sigma$, i.e., $E(S(w))=\sigma$.\\
\item[{\rm (ii)}] The mean squared error {\rm (MSE)} of $S(w)$ is \\
${\rm MSE}(S(w)) = {\rm Var}\left[w(\dfrac{b-a}{\xi})\right]+{\rm Var}\left[(1-w)(\dfrac{q_{3}-q_{1}}{\eta})\right]+2{\rm Cov}\left[w(\dfrac{b-a}{\xi}),(1-w)(\dfrac{q_{3}-q_{1}}{\eta})\right]$.\\
\item[{\rm (iii)}] The optimal weight is\\
${w_{\rm opt}}=\dfrac{{\rm Var}(q_{3}-q_{1})/\eta^{2}-{\rm Cov}(b-a, q_{3}-q_{1})/(\xi\eta)}{{\rm Var}(b-a)/\xi^{2}+{\rm Var}(q_{3}-q_{1})/\eta^{2}-2{\rm Cov}(b-a, q_{3}-q_{1})/(\xi\eta)}$.
\end{enumerate}
\end{theorem}

\vskip 12pt
\noindent{\bf Proof}.
(i) The expected value of the proposed estimator is
$$E(S(w))=wE\left(\dfrac{b-a}{\xi}\right)+(1-w)E\left(\dfrac{q_{3}-q_{1}}{\eta}\right).$$
Noting that $E\left((b-a)/\xi\right)=\sigma$ and $E\left((q_{3}-q_{1})/\eta\right)=\sigma$, we have $E(S(w))=\sigma$.

\vskip 12pt
(ii) By part (i), we have Bias($S(w)$) = 0. Then
\begin{equation}\nonumber
\begin{split}
 {\rm MSE}(S(w))
  = \ &{\rm Var}(S(w))\\
  = \ &{\rm Var}\left[w(\dfrac{b-a}{\xi})\right]+{\rm Var}\left[(1-w)(\dfrac{q_{3}-q_{1}}{\eta})\right]+2{\rm Cov}\left[w(\dfrac{b-a}{\xi}),(1-w)
                       (\dfrac{q_{3}-q_{1}}{\eta})\right]  \\
  = \ & \left[{\rm Var}\left(\dfrac{b-a}{\xi}\right)+{\rm Var}\left(\dfrac{q_{3}-q_{1}}{\eta}\right)-2{\rm Cov}\left(\dfrac{b-a}{\xi},\dfrac{q_{3}-q_{1}}{\eta}\right)\right]
   w^{2}\\
   & -\left[2{\rm Var}\left(\dfrac{q_{3}-q_{1}}{\eta}\right)-2{\rm Cov}\left(\dfrac{b-a}{\xi},\dfrac{q_{3}-q_{1}}{\eta}\right)\right]w+{\rm Var}\left(\dfrac{q_{3}-q_{1}}{\eta}\right).
\end{split}
\end{equation}

(iii) The first derivative of MSE with respect to $w$ is
\begin{align*}
  {d\over dw} {\rm MSE}(S(w))= \ & 2\left[{\rm Var}\left(\dfrac{b-a}{\xi}\right)+{\rm Var}\left(\dfrac{q_{3}-q_{1}}{\eta}\right)-2{\rm Cov}\left(\dfrac{b-a}{\xi},\dfrac{q_{3}-q_{1}}{\eta}\right)\right]w \\
            & -\left[2{\rm Var}\left(\dfrac{q_{3}-q_{1}}{\eta}\right)-2{\rm Cov}\left(\dfrac{b-a}{\xi},\dfrac{q_{3}-q_{1}}{\eta}\right)\right].
\end{align*}
By setting the first derivative equal to zero, we have the solution as
\begin{equation}\label{solution}
w=\dfrac{{\rm Var}(q_{3}-q_{1})/\eta^{2}-{\rm Cov}(b-a,q_{3}-q_{1})/(\xi\eta)}{{\rm Var}(b-a)/\xi^{2}+{\rm Var}(q_{3}-q_{1})/\eta^{2}-2{\rm Cov}(b-a,q_{3}-q_{1})/(\xi\eta)}.
\end{equation}
Note also that, by the Cauchy-Schwartz inequality,\\
$${d^2\over dw^2} {\rm MSE}(S(w))=2\left[{\rm Var}\left(\dfrac{b-a}{\xi}\right)+{\rm Var}\left(\dfrac{q_{3}-q_{1}}{\eta}\right)-2{\rm Cov}\left(\dfrac{b-a}{\xi},\dfrac{q_{3}-q_{1}}{\eta}\right)\right]\geq0.$$
This shows that ${\rm MSE}(S(w))$ is a convex function of $w$, and hence the solution in (\ref{solution}) is the optimal weight associated with the proposed estimator.

\end{appendices}

\end{document}